\documentclass[letterpaper,11pt]{article}
\pdfoutput=1
\usepackage{jheppub}
\usepackage[framemethod=TikZ]{mdframed}
\usepackage{amsmath,amssymb,amsfonts,bm,amscd,mathtools}
\usepackage{graphicx}
\usepackage{multirow}
\usepackage{verbatim}
\usepackage{appendix}
\usepackage{fancybox}
\usepackage{array}
\usepackage{url}
\usepackage{float}
\usepackage{subfigure}

\newcounter{ex}[section]
\renewcommand{\theex}{\arabic{section}.\arabic{ex}}
\newenvironment{ex}[2][]{%
\refstepcounter{ex}%
\ifstrempty{#1}%
{\mdfsetup{%
frametitle={%
\tikz[baseline=(current bounding box.east),outer sep=0pt]
\node[anchor=east,rectangle,fill=gray!20]
{\strut Exercise~\theex};}}
}%
{\mdfsetup{%
frametitle={%
\tikz[baseline=(current bounding box.east),outer sep=0pt]
\node[anchor=east,rectangle,fill=gray!20]
{\strut Exercise~\theex:~#1};}}%
}%
\mdfsetup{innertopmargin=10pt,linecolor=gray!20,%
linewidth=2pt,topline=true,%
frametitleaboveskip=\dimexpr-\ht\strutbox\relax
}
\begin{mdframed}[]\relax%
\label{#2}}{\end{mdframed}}

\newcounter{example}[section]
\renewcommand{\theexample}{\arabic{section}.\arabic{example}}

\newcounter{fact}[section]
\renewcommand{\thefact}{\arabic{section}.\arabic{fact}}

\newcommand{\Q}{{\mathcal Q}}

\def\Tr{{\rm Tr \,}}

\def\CI{{\mathcal I}}

\def\CM{{\mathcal M}}
\def\CN{{\mathcal N}}
\def\CO{{\mathcal O}}

\def\CQ{{\mathcal Q}}

\def\CS{{\mathcal S}}

\def\II{{\mathcal I}}
\def\a{\alpha}

\def\be{\begin{equation}}
\def\ee{\end{equation}}
\def\bea{\begin{eqnarray}}
\def\eea{\end{eqnarray}}

\title{Lectures on the Superconformal Index}

\author{Abhijit Gadde}
\affiliation{Department of Theoretical Physics \\ 
 Tata Institute for Fundamental Research, Mumbai 400005
}

\abstract{
In these lectures, we give a pedagogical introduction to the superconformal index.  This is the writeup of the lectures given at the Winter School ``YRISW 2020" and is to appear in a special issue of JPhysA. The lectures are at a basic level and are geared towards a beginning graduate student interested in working with the superconformal index.
}

\preprint{TIFR/TH/20-20}

\begin{document}
\maketitle
\flushbottom

\section{Plan of lectures}
The superconformal index is the Witten index for superconformal field theories in radial quantization. We will first introduce the Witten index in the context of supersymmetric quantum mechanics in section \ref{witten-index}. In addition to understanding basic principles behind the protection on the index, we develop many tools, in particular plethystic exponentiation, that would be necessary to compute the superconformal index of gauge theories in a simpler setting.  In section \ref{sci}, we consider the superconformal algebra and introduce the superconformal index. The index of basic ingredients i.e. the chiral multiplet and the vector multiplet is computed. Then we discuss the invariance of the index under recombination of short multiplets of superconformal algebra. We conclude that the superconformal index captures all the spectral information of the theory up to recombinations. In section \ref{n=2index} we apply the superconformal index to the so called ``class $\CS$" dualities. We describe the topological field theory associated with the index after explaining limits of the index with enhanced supersymmetry. Lastly, in section \ref{largen}, we compute the index of $\CN=4$ $SU(N)$ super Yang-Mills in the large $\CN$ limit. A priori, supersymmetric black-holes in $AdS_5$ are supposed to contribute to the index $\CO(N^2)$ states. But we find that the index is $\CO(1)$ and is saturated only by the supergravity states in the bulk. We comment on this apparent mismatch and describe some recent work on recovering the black-hole entropy from the index. Some other aspects of the index, such as its modular property \cite{Razamat:2012uv, Gadde:2020bov} and its connection to anomaly \cite{Spiridonov:2012ww}, will not be discussed in these lectures. We will only discuss the index on $S^3$. For discussion of the index on lens spaces, see  \cite{Benini:2011nc, Razamat:2013opa, Razamat:2013jxa, Kels:2017toi}. For some other excellent reviews on the superconformal index, see \cite{Rastelli:2016tbz, Rastelli:2014jja}.

\section{Witten index}\label{witten-index}

\subsection{Supersymmetric harmonic oscillator}
Consider a supersymmetric harmonic oscillator
\be
L=\frac12 {\dot x}^2-\frac12   x^2+i {\bar \psi} {\dot \psi}- {\bar \psi}{\psi}.
\ee
The Hamiltonian of the system is
\bea
H&=&{\dot \psi} {\Pi_\psi}+{\dot {\bar \psi}} {\Pi_{\bar \psi}}+{\dot x}p-L\\
&=& \frac12 p^2+\frac12  x^2 + {\bar \psi}{\psi}.
\eea 
where $\Pi_\psi, \Pi_{\bar \psi}$ and $p$ are canonical conjugates to $\psi,{\bar \psi}$ and $x$ respectively. Quantizing the system leads to creation and annihilation operators 
\be
a_B^{\dagger}=\frac{1}{\sqrt{2}}(-ip+ x), \qquad a_B=\frac{1}{\sqrt{2}}(ip+ x),\qquad a_F^\dagger={\bar \psi},\qquad a_F=\psi,
\ee
which obey $[a_B,a_B^\dagger]=1$ and $\{a_F,a_F^\dagger\}=1$.
The Hamiltonian in terms of these operators is
\be\label{hamilt}
H=( a_B^\dagger a_B+a_F^\dagger a_F).
\ee
It is not difficult to see that this Hamiltonian has a supersymmetry $Q=a_B^\dagger a_F$ and ${\bar Q}= a_F^\dagger a_B$. The algebra of these charges is
\be
[Q,H]=[{\bar Q},H]=0,\qquad \{Q,{\bar Q}\}=H.
\ee
This is a universal feature of the supersymmetry algebra. In addition to these continuous symmetries, the theory also has a discrete ${\mathbb Z}_2$ symmetry that is relevant for our discussion, namely the fermion number $F$. The operator $F$ counts the number of fermions in the state modulo $2$. 

It is easy to compute the spectrum.  As the ground state obeys $H|0\rangle =0$, it also obeys $Q|0\rangle = 0$. Excited states are created by acting on the vacuum by creation operators. 
\be
|\chi_B,n\rangle \equiv  (a_B^\dagger)^n |0\rangle, \qquad |\chi_F,m\rangle\equiv (a_B^\dagger)^m \,a_F^\dagger\, |0\rangle.
\ee
The first set of states have $F=0$ and the second set of states have $F=1$.
The partition function is also computed easily
\be
Z(x)={\Tr}\, x^{H}= 1+(x+x^{2}+\ldots )+(x+x^{2}+\ldots )=\frac{1+ x}{1-x},\qquad x=e^{-\beta}.
\ee

Let us explain in detail, a trick that can be used to compute partition function of the free theory. As explained above, the spectrum is generated simply by the action of all the creation operators on the vacuum. These creation operators commute and hence act independently. It is convenient to compute the partition function over ``single letters" i.e. on the states with only a single particle. This is called the single letter partition function and is denoted as $z(x)$. It is useful to compute it separately for bosons and fermions. They are denoted by $z_B$ and $z_F$. In this case, there is only one bosonic creation operator  and one fermionic creation operator. 
\bea
z(x)&=&z_B(x)+z_{F}(x)\nonumber\\
z_B(x)&=&{\rm Tr}_{\text{bosonic letters}} x^H= x\nonumber\\
z_F(x)&=&{\rm Tr}_{\text{fermionic letters}} x^H= x.
\eea

The partition function over bosonic multi-particle states is obtained by
\be
z_B(x)=x\to \frac{1}{1-x}\equiv Z_B(x),\qquad z_F(x)= x \to (1+ x)\equiv Z_F(x).
\ee
And the full partition function is $Z=Z_B Z_F$. For bosons $Z_B$ corresponds to sum over all the states obtained by acting with any number of creation operators while for fermions $Z_F$ corresponds to sum over state with no fermion and a single fermion.
If the theory has multiple types of creation operators, the single letter partition function would have more terms. In order to compute the full partition function we would like to do this replacement for every monomial in $z_B$ and $z_F$ and then take the  product. This is achieved by a mathematical operation called ``plethystic exponentiation" i.e. PE. 

We define 
\be
PE[f(x_i)]=\exp\Big(\sum_{n=1}^\infty\frac1n f(x_i^n)\Big),\qquad {\widetilde {PE}}[f(x_i)]=\exp\Big(-\sum_{n=1}^\infty\frac1n (-1)^n f(x_i^n)\Big)
\ee 
Then,
\be
\boxed{
Z_B(x)=PE[z_B(x)],\qquad Z_F(x,{\tt f})={\widetilde {PE}}[z_F(x,{\tt f})],\qquad Z=Z_B Z_F.}
\ee

\subsection*{The Index}
In a theory with interactions, we will not be able to compute the partition function it so easily. 
The Lagrangian of a general interacting theory that preserves the supersymmetry is
\be
L=\frac12 {\dot x}^2-\frac12 W'(x)^2+i {\bar \psi} {\dot \psi}- W''(x) {\bar \psi}{\psi}.
\ee
Here the function $W(x)$ is called superpotential. For $W(x)= x^2/2$ we get back the supersymmetric simple harmonic oscillator. It is clear that for any higher polynomial  $W(x)$, the theory will be interacting. The spectrum of the theory, and hence the partition function,  is difficult to compute exactly. Nevertheless there is a ``protected quantity" that can be computed exactly. 

Let us observe that all the states with energy $E\neq 0$ are degenerate with degeneracy two. One state is bosonic and the other fermionic. This is simply a consequence of supersymmetry.  Let $|\psi\rangle$ be an eigenstate of the Hamiltonian with energy $E\neq 0$, 
\be
H({\bar Q} |\psi\rangle)={\bar Q} H |\psi\rangle= E({\bar Q} |\psi\rangle).
\ee
The spectrum of states \emph{apart from the zero energy state} is paired. This pairing of non-zero energy states is a robust phenomenon having only to do with supersymmetry and is expected even when the theory has interactions that preserve supersymmetry. As we turn on the interactions, the spectrum of the theory will change but  because of supersymmetry, the states will always respect pairing.  States can become zero energy but in pairs. States can leave zero energy but in pairs. This observation lead Witten to define an index \cite{Witten:1982df}, 
\be
I={\rm Tr} \, (-1)^F \, x^{ H}.
\ee
It is seems like it is a small tweak over the partition function but it makes a big difference. The factor of $(-1)^F$ leads to cancellations of bosonic and fermionic contributions at non zero energy and gets contribution only from zero energy states. In the case of the supersymmetric harmonic oscillator, explicitly
\be
I={\Tr}\,(-1)^F x^{ H}= 1+(x^{ }+x^{2 }+\ldots )-(x^{ }+x^{2 }+\ldots )=1.
\ee
Above arguments tell us that even if we turn on supersymmetry preserving interactions in this theory, the index will always remain $1$\footnote{One should make sure that the deformation of the theory does not lead to a non-normalizable vacuum state because in such transitions new zero energy states could come from infinity. See \cite{Cooper:1994eh} for more discussion in the context of supersymmetric quantum mechanics.}. In other words, the index of the interacting theory can be computed by setting the interactions strength to zero as the index is invariant under smooth deformations. In this way, the plethystic techniques developed before become useful to compute the index.

The index only receives contributions from states with $H=0$ and as energy is additive in a free theory, the index can be constructed starting from ``single letter index" as well.
To compute the partition function of the free theory we had two different plethystic formulas for bosons and fermions. There is even a simpler way to construct multi-particle index from a single particle index
\be
\boxed{
I=PE[i(x)],\qquad i(x)=i_B(x)-i_F(x).
}
\ee
Here,
\bea
i_B(x)&=&{\rm Tr}_{\text{(bosonic letters with $H=0$)}} x^H\nonumber\\
i_F(x)&=&{\rm Tr}_{\text{(fermionic letters with $H=0$)}} x^H.
\eea
\begin{ex}[]{wittenpe}
Show this.
\end{ex}
In the case of the supersymmetric harmonic oscillator, the bosonic single letter index $i_B(x)=0$ and fermionic single letter index $i_F(x)=0$. Hence $i(x)=0$. This gives the full index to be
\be
I(x)=PE[i(x)]=1.
\ee

To reiterate, for an interacting supersymmetric theory, the partition function is non-trivial to compute, in particular it is a function of $x$ while the index is easy to compute (by setting coupling to zero) and is independent of $x$ (because only the states with $H=0$ contribute). 

\subsection{Global symmetry}

Consider $2N$ copies of the above Lagrangian
\be
L=\frac12 |{\dot x_i}|^2-\frac12   |x_i|^2+i {\bar \psi_i} {\dot \psi_i}- {\bar \psi_i}{\psi_i}.
\ee
This has $SO(2N)$ symmetry. There is a way to introduce interactions that preserve this symmetry,
\be
L=\frac12 {\dot x_i}^2-\frac12 |\frac{\partial W(x_i)}{\partial x_i}|^2+i {\bar \psi_i} {\dot \psi_i}-\frac{\partial^2 W(x)}{\partial x_i \partial x_j} {\bar \psi_i}{\psi_j}.
\ee
For $W= |x_i|^2$, we get the free theory. Let us focus on the free theory for now. Because the theory is free, we can compute the partition function easily. It consists $2N$ bosonic creation operators  $a_{B,i}^\dagger$ and $2N$ fermionic creation operators $a_{F,i}^\dagger$. 

This means 
\be
z_B(x)=2N x,\qquad z_F(x)=2N x.
\ee
Using the plethystic formula,
\be
Z(x)=\Big(\frac{1+x^{ }}{1-x^{ }}\Big)^{2N}.
\ee

Sometimes it is convenient to keep track of representation of states under global symmetry. In this case, we have $SO(2N)$ symmetry. This is done by turning on fugacities with respect to the Cartan generators.
\be
Z(x,a_i)={\rm Tr}\, x^H\, a_i^{J_i},\qquad i=1,\ldots N.
\ee
It is best to compute this using single letter partition function. As the bosons transform in fundamental representation of $SO(N)$. Their single letter partition function is
\be
z_B(x,a_i)=x\Big(a_1+\frac{1}{a_1}+\ldots + a_n+\frac{1}{a_n}\Big)=x \,\chi_{fund}(a_i)
\ee 
Here $\chi_{fund}(a_i)$ is the character of the fundamental representation. In general, the character of a representation is defined as
\be
\chi_R(a_i)=\sum_{\rho\in R} \prod_i\, a_i^{\rho_i}\equiv  \sum_{\rho\in R} a^\rho
\ee
where $\rho=(\rho_1,\ldots, \rho_N)$ is a weight vector and the sum is taken over all the weight vectors of the representation. Sometimes we will abbreviate $\prod_i\, a_i^{\rho_i}=a^\rho$. Recall that the non-zero weight vectors of the adjoint representation are called roots. So we have,
\bea
z_B(x,a_i)&=&x\Big(a_1+\frac{1}{a_1}+\ldots + a_n+\frac{1}{a_n}\Big)=x \,\chi_{fund}(a_i)\nonumber\\
Z_B(x,a_i)&=& PE[z(x,a)]= \prod_{i=1}^N\frac{1}{(1-x a_i) (1-x /a_i)}\nonumber\\
z_F(x,a_i)&=&x\Big(a_1+\frac{1}{a_1}+\ldots + a_n+\frac{1}{a_n}\Big)= x\, \chi_{fund}(a_i)\nonumber\\
Z_F(x,a_i)&=& {\widetilde {PE}}[z(x,a)]= \prod_{i=1}^N (1+x a_i) (1+x /a_i)\nonumber\\
Z(x,a_i)&=& Z_B(x,a_i) Z_F(x,a_i)=\prod_{i=1}^N \frac{(1+x a_i) (1+x /a_i)}{(1-x a_i) (1-x /a_i)}\equiv \prod_{\rho} \frac{(1+x a^\rho) }{(1-x a^\rho)}.
\eea
It is clear that when we set $a_i=1$, then $Z(x,a_i)$ reduces to the unrefined partition function $Z(x)$.
The advantage of keeping track of charges with respect to the Cartan generators is that we can compute partition function over states with a given representation of global symmetry. This uses the orthogonality of characters,
\be
\frac{1}{|W|}\oint \prod_{i=1}^N \frac{d a_i}{2\pi i a_i}\,  \Delta(a_i) \chi_R(a_i)\, \chi_{R'}(a_i)=\delta_{RR'}.
\ee
where $N$ is the rank of the group, $|W|$ is the cardinality of the associated Weyl group and  $\Delta(a_i)$ is called the Van-der-Monde determinant. The van-der-Monde determinant has a universal formula in terms of plethystic exponent\footnote{Requires regularization for the plethystic exponent to be well-defined.}
\be
\Delta(a_i)=(PE[\sum_\alpha a^{\alpha}])^{-1}=\prod_\alpha (1-a^\alpha),
\ee
where $\alpha=(\alpha_1,\ldots, \alpha_N)$ are the roots of the Lie algebra.
Now it is clear how to project onto states with given representation,
\be
Z(x,a_i)|_R=\frac{1}{|W|}\oint \prod_{i=1}^N \frac{d a_i}{2\pi i a_i}\Delta(a_i) Z(x,a_i) \chi_R(a_i).
\ee
In computing the index of gauge theories, we will need to project on states with trivial representation. The trivial representation has the character  $\chi_R(a_i)=1$. 

\begin{ex}[]{ex1}
Perform this integral in $x$ expansion and also using Cauchy's theorem for $SO(2)$.
\end{ex}

Again, the single letter index is $i_B(x,a)=0$ and $i_F(x,a)=0$. This implies $I(x,a)=PE[i_B(x)-i_F(x)]=1$. Note that in the index only the fugacities for those charges that commute with the supercharge can be turned on. This is because the robustness of the index relies on the Bose-Fermi cancellations for non-zero energy states. If a fugacity for the charge that doesn't commute with $Q$ is turned on then the cancellation will not go through and the index will not be protected under quantum corrections.

So far we have not encountered a theory with a non-trivial index but we have developed tools that would help us in computing the index when it is non-trivial. Now we will turn our attention to superconformal index.

\section{The superconformal index}\label{sci}
The  index of a $4d$ superconformal field theory is defined
as the Witten index of the theory in radial quantization.  Let ${\cal Q}\equiv \CQ_-$ be  one of the  Poincar\'e supercharges, and $\Q^\dagger ={\cal  S}_+$
the conjugate conformal supercharge. Schematically, the index is defined as ~\cite{Kinney:2005ej, Romelsberger:2005eg, Romelsberger:2007ec}
 \be \label{basicdef} {\mathcal I} (\mu_i)={\rm
Tr}\, (-1)^{F}\,x^{\delta}\, \mu_i^{ {\cal M}_i}\, ,
\ee
where the trace is over the Hilbert space  of the theory on $S^3$,
$\delta \equiv \frac{1}{2} \{\Q,\, \Q^{\dagger}\}$. Since states with $\delta >0$
come in boson/fermion pairs, only the $\delta =0$ states contribute, and the index is independent of $x$. As remarked earlier, the Fermi-Bose cancellation works only if the additional fugacities $\mu_i$ are turned on only for those charges $\CM_i$ that commute with $\CQ$. 
Unlike in the case of quantum mechanics, this theory has  infinitely many states with $\delta =0$. Not just that, as will see shortly, this theory has infinitely many ``single letters" with $\delta=0$.
The introduction of the fugacities $\mu_i$ serves both to regulate this divergence and to achieve a more
refined counting.

For  ${\cal N}=1$, the supercharges are $\{ {\cal Q}_\alpha \, , {\cal S}^\alpha \equiv {\cal Q}^{\dagger\, \alpha} \, , {\widetilde {\cal Q}_{\dot \alpha}} \, ,
\widetilde {\cal S}^{\dot \alpha} \equiv {\widetilde {\cal Q}^{\dagger \,\dot \alpha}} \}$,
where $\alpha = \pm$ and $\dot \alpha = \dot \pm$ are respectively $SU(2)_1$ and $SU(2)_2$ indices, with  $SU(2)_1 \times SU(2)_2 = Spin(4)$ the isometry group of the $S^3$.
The relevant anticommutators are
\bea\label{N1Q}
\{\Q_\alpha, \, {\cal \Q}^{\dagger\, \beta} \} & =& \Delta+2M_\alpha^\beta+\frac{3}{2}r \\
\{\widetilde \Q_{\dot \alpha}\,, \, {\widetilde {\cal \Q}}^{\dagger \, \dot \beta} \} & =& \Delta +2 \widetilde M_{\dot \alpha}^{\dot \beta}-\frac{3}{2}r \, ,
\eea
where $\Delta$ is the conformal dimension, $M_{\alpha}^\beta$ and  $\widetilde M_{\dot \alpha}^{\dot \beta}$  the $SU(2)_1$ and $SU(2)_2$ generators, and
$r$ the generator of the $U(1)_r$ R-symmetry. In our conventions, the $\Q$s have $r=-1$ and $\widetilde \Q$s have $r=+1$,
and of course the dagger operation  flips the sign of $r$. 

With the choice of supercharge $\CQ\equiv \CQ_-$ to define the index,   we get $\delta=\Delta-2j_1+\frac{3}{2}r$. The commutant of $\CQ$ in the superconformal algebra $SU(2,2|\CN)$ is $SU(2,1|\CN-1)$. For $\CN=1$, the commutant  is $SU(2,1)$.  It has rank $2$. We can choose an arbitrary basis of charges in this space as charges $\CM_i$ to refine the index. 
In order to agree with the existing literature, we choose it to be $\frac13(\Delta+j_1)\pm j_2$. The index is then defined as
\be
\II(p,q) \equiv {\rm Tr} \, (-1)^F p^{\frac13(\Delta+j_1)+j_2} q^{\frac13(\Delta+j_1)-j_2} =   {\rm Tr} \, (-1)^F p^{j_1-\frac12 r+j_2}q^{j_1-\frac12 r-j_2}\, ,\quad
\delta=\Delta-2j_1+\frac{3}{2}r \, ,
\ee
In the second equality we have replaced the charge $\frac13(\Delta+j_1)$ by $j_1-\frac12 r$ using the fact that only states with $\delta=0$ contribute to the index. This is done because, the r-charges are easier to determine than conformal dimensions. 

\subsection{Chiral multiplet}
Let us start with the theory of a single free  chiral multiplet $\Phi$. It obeys the condition ${\bar \CQ}_{\dot \alpha} \Phi=0$. The superspace expansion of the chiral multiplet is
\be
\Phi = \phi+\sqrt{2} \theta \psi + i\theta^\dagger {\bar \sigma}^\mu \theta \partial_\mu \phi
\ee
The complex conjugate is an anti-chiral field i.e. $\CQ_\alpha {\bar \Phi}=0$. The r-charge of $\phi$ is $2/3$ (do you know why?).

In radial quantization, the Hilbert space of the theory is isomorphic to the spectrum of local operators.  These are constructed using fields from the chiral multiplet and their derivatives. This means the analogue of our creation operators in quantum mechanics are fields and their derivatives. They are
\be
\phi, \partial_\mu \phi, \partial_\mu\partial_\nu \phi,\ldots, \qquad \psi, \partial_\mu \psi, \partial_\mu \partial_\nu \psi,\ldots
\ee
and their conjugates. Here the derivatives act with all possible contractions but with the constraint $\partial_\mu \partial^\mu \phi=0$ that comes from the equation of motion. Similarly, the fermion also obeys the equation of motion $\partial_{\alpha{\dot \alpha}}\psi^\alpha=0$. With these constraints we  tabulate the letters and enumerate the ``single letter index" in table \ref{tab1}.

\begin{table}[htbp]
\begin{centering}
\begin{tabular}{|c||c|c|c|c||c|c|}
\hline Letters & $\Delta$ & $j_{1}$ & $j_{2}$ & $r$ 
& $\delta$ & ${\cal I}$  \tabularnewline \hline \hline 
$\phi$ &
$1$ & $0$ & $0$ & $\frac{2}{3}$ &  $2$ & $$ \tabularnewline \hline 
$\psi_+$ & $\frac{3}{2}$ &
$\frac{1}{2}$ & $0$ & $-\frac{1}{3}$  & $0$ &
$-(pq)^{\frac23}$  \tabularnewline \hline 
$\psi_-$ & $\frac{3}{2}$ &
$-\frac{1}{2}$ & $0$ & $-\frac{1}{3}$  & $2$ &
$$  \tabularnewline \hline 
$\partial^\alpha_{\dot +} \psi_\alpha$ &
$\frac{5}{2}$ & $0$ & $\frac{1}{2}$ & $-\frac{1}{3}$ &
$2$ & $$
\tabularnewline \hline 
$\partial^\alpha_{\dot -} \psi_\alpha$ &
$\frac{5}{2}$ & $0$ & $-\frac{1}{2}$ & $-\frac{1}{3}$ &
$2$ & $$
\tabularnewline \hline 
 $\square\phi$ & $3$ & $0$ & $0$ &
$\frac{2}{3}$  & $4$ & $$ \tabularnewline \hline
\hline 
$\bar{\phi}$ & $1$ & $0$ & $0$ & $-\frac{2}{3}$ & $0$
& $(pq)^{\frac13}$ \tabularnewline \hline 
$\bar{\psi}_{\dot +}$ &
$\frac{3}{2}$ & $0$ & $\frac{1}{2}$ & $\frac{1}{3}$ & 
$2$ & $$ \tabularnewline \hline
$\bar{\psi}_{\dot -}$ &
$\frac{3}{2}$ & $0$ & $-\frac{1}{2}$ & $\frac{1}{3}$ & 
$2$ & $$ \tabularnewline \hline
$\partial^{\dot \alpha}_{+}\bar{\psi}_{\dot \alpha}$ & $\frac{5}{2}$ & $\frac{1}{2}$ & $0$ &
$\frac{1}{3}$ & $2$ & $$  \tabularnewline
\hline 
$\partial^{\dot \alpha}_{-}\bar{\psi}_{\dot \alpha}$ & $\frac{5}{2}$ & $-\frac{1}{2}$ & $0$ &
$\frac{1}{3}$ & $4$ & $$  \tabularnewline
\hline 
$\square\bar{\phi}$ & $3$ & $0$ & $0$ & $-\frac{2}{3}$ 
& $2$ & $$
 \tabularnewline \hline\hline
$\partial_{+{\dot +}}$ & $1$ & $\frac{1}{2}$ & $\frac{1}{2}$ &
$0$ &  $0$ & $p$ \tabularnewline \hline
$\partial_{-{\dot +}}$ & $1$ & $-\frac{1}{2}$ & $\frac{1}{2}$ &
$0$ &  $0$ & $q$ \tabularnewline \hline
$\partial_{+{\dot -}}$ & $1$ & $\frac{1}{2}$ & $-\frac{1}{2}$ &
$0$ &  $2$ & $$ \tabularnewline \hline
$\partial_{-{\dot -}}$ & $1$ & $-\frac{1}{2}$ & $-\frac{1}{2}$ &
$0$ &  $2$ & $$ \tabularnewline \hline
\end{tabular}
\par\end{centering}
\caption{\label{tab1}The ``letters'' of an ${\cal N}=1$ chiral
multiplet and their contributions to the index. Here $\delta=\Delta-2j_{1}+\frac{3}{2}r$.  A priori
we have to take into account the free equations of motion $\partial \psi =0$ and $\Box \phi =0$,
which imply constraints on the possible words,
but we see that in this case equations of motions have $\delta \neq 0$ so they do not contribute to the single letter index. Finally there
are two spacetime derivatives contributing to the index (i.e. have $\delta=0$), and their tower on the fields is responsible for the denominator of the single letter index.
}
\end{table}
From here, we can easily compute the single letter index,
\be
i_\phi(p,q)=\frac{(pq)^{\frac13}-(pq)^{\frac23}}{(1-p)(1-q)}.
\ee
The denominators encode the action of the two spacetime derivatives with $\delta = 0$.
If the chiral multiplet transforms in representation $R$ of some global symmetry. Then we can refine the single letter index refined with the fugacities for the Cartan generators. Note that if the chiral multiplet transforms in representation $R$ then the anti-chiral multiplet transforms in the conjugate representation $\bar R$.
\be
i_\phi(a_i;p,q)=\frac{(pq)^{\frac13}\, \chi_{\bar R}(a_i)-(pq)^{\frac23} \, \chi_{R}(a_i)}{(1-p)(1-q)}.
\ee

\begin{ex}[]{ex2}
Define the index with respect to ${\widetilde{\CQ}}_{\dot +}$ and compute the single letter index of $\Phi$ by making a similar table.
\end{ex}

The full index is obtained by plethystic exponentiation. For a single multiplet charged $1$ under a $U(1)$ symmetry, it is
\be
\CI_\phi(a;p,q)=\Gamma((pq)^{\frac13} a^{-1};p,q),\qquad \quad \Gamma(z;p,q)\equiv \prod_{i,j=1}^\infty\frac{1-p^{i+1}q^{j+1}/z}{1-p^iq^j z}.
\ee
The special function $\Gamma(z;p,q)$ is known as the elliptic gamma function \cite{Felder_2000} and it studied in the math literature as the ``elliptic" generalization of the usual Gamma function. The elliptic gamma function plays an important role in the theory of superconformal indices because chiral multiplets are an integral part of any superconformal Lagrangian in four dimensions. We will almost always drop the arguments $(p,q)$ of the elliptic Gamma function and denote $\Gamma(z;p,q)\to \Gamma(z)$.
If the chiral multiplet transforms in a representation $R$ then the corresponding index is
\be
\CI_\phi(a_i;p,q)=\prod_{\rho\in {\bar R}} \Gamma((pq)^{\frac13} a^\rho).
\ee

If a chiral multiplet participates in a superpotential interaction then its R-charge can be different from $2/3$. It is determined by demanding the superpotential has R-charge $2$. This constraint does not fix the R-charges completely if the theory has abelian global symmetry as they could mix with this putative R-charge. In that case, the R-charge is determined by insisting that the R-symmetry does not have mixed anomaly with any of the abelian global symmetries. This procedure is known as ``a-maximization" \cite{Intriligator:2003jj, Kutasov:2003iy}. The single letter index of the chiral multiplet with R-charge $r$ is computed by summing over  letters that contribute to the index in the free case but they contribute to the index according to their R-charge. This is known as Romelsberger's prescription \cite{Romelsberger:2005eg, Romelsberger:2007ec}. This peculiar prescription was later understood by preserving supersymmetry through out the RG flow on $S^3\times S^1$ \cite{Festuccia:2011ws}.

From table \ref{tab1}, the letters of the chiral multiplets that contribute to the index are $\psi_+$ and $\bar \phi$. If the chiral multiplet has R-charge $r$, then the R-charges of these two letters will be $r-1$ and $-r$ respectively. Accordingly they will contribute $-(pq)^{1-\frac12 r}$ and $(pq)^{\frac12 r}$ respectively. This gives the index of the chiral multiplet with R-charge $r$ to be
\be
i_{\phi_r}(p,q)=\frac{(pq)^{\frac12 r}-(pq)^{1-\frac12 r}}{(1-p)(1-q)}.
\ee
The multi-particle index is
\be
\CI_{\phi_r}(a;p,q)=\Gamma((pq)^{\frac12 r}a).
\ee

\subsection*{Examples}
With these tools we can already compute the index of some physical theories involving only chiral multiplets.  Examples in this subsection are taken from the review \cite{Rastelli:2016tbz}.
\subsection* {Massive theory: } 
Consider the theory of a single chiral multiplet with a mass term i.e. $W(\Phi)=m \Phi^2$. This fixes the R-charge of $\phi$ to be $1$. 
\be
\CI_{\phi_1}(p,q)=\Gamma((pq)^\frac12 )=1.
\ee
This is consistent with the physical expectation as the massive theory should have a single supersymmetric ground state.

One can consider as a slight generalization a theory of two chiral multiplets with the mass type superpotential $W(\Phi_i)=m \Phi_1\Phi_2$. This superpotential is not sufficient to fix the R-charges of $\Phi_{1,2}$ completely but it does tell us that $r_1+r_2=2$. Just with this constraint
\be
\CI(a;p,q)=\Gamma((pq)^{\frac12 r} a)\Gamma((pq)^{1-\frac12 r} a^{-1})=1.
\ee
This is also to be expected because this superpotential gaps out both chiral multiplets as is clear from the Lagrangian written in components.

\subsection*{F-term supersymmetry breaking}
Consider a simple O'Raifeartaigh type model with a linear superpotential $W(\Phi)=\eta \Phi$. This theory breaks supersymmetry spontaneously as the this superpotential imparts non-zero energy to the ground state. Let us see how this works for the index. The superpotential fixes the R-charge of $\Phi$ to be $2$ and does not allow for any global symmetry.
\be
\CI_{\phi_2}(p,q)=\Gamma(pq)=0.
\ee
The zero in the index is coming because the contribution of $\psi_+$ becomes $-(pq)^{1-\frac12 r}\to -1$. This means that the state with $\psi_+|0\rangle$ has the same energy as $|0\rangle$. In other words, the ground state comes in a bose-fermi pair and hence is not protected. This also means $\psi_+$ should be interpreted as Goldstino mode  for spontaneous breaking of supersymmetry. Any O'Raifeartaigh type model that breaks supersymmetry spontaneously always has a neutral chiral multiplet of charge $2$. This results in $\CI=0$.

\subsection*{Non-trivial chiral ring}
A simplest model with a non-trivial chiral ring is a theory of a single chiral multiplet with superpotential $W(\Phi)=\Phi^{n+1}$. The chiral ring relation is $W'(\Phi)=0=\Phi^n$. The superpotential means that field $\Phi$ has R-charge $2/(n+1)$ and also that it is charged $1$ with respect to a discrete symmetry ${
\mathbb Z}_{n+1}$. We can compute the index with a fugacity $g$ to the flavor symmetry turned on. As the flavor symmetry is only discrete ${\mathbb Z}_{n+1}$, the fugacity obeys $g^{n+1}=1$.
The single letter index is
\be
i(g;p,q)=\frac{(pq)^{\frac{1}{n+1}}g-(pq)^{1-\frac{1}{n+1}}g^{-1}}{(1-p)(1-q)}=\frac{(pq)^{\frac{1}{n+1}}g-((pq)^{\frac{1}{n+1}}g)^{n}}{(1-p)(1-q)}.
\ee
The contribution coming from the fermionic letter $\psi_+$ is  exactly n-times the contribution of the bosonic letter and with opposite sign. This is in accordance with the chiral ring relation $\Phi^n=0$.

\subsection{Index of gauge theory}

In addition to the $\CN=1$ chiral multiplet, a ubiquitous multiplet in 4d superconformal theories is the $\CN=1$ vector multiplet. The index of the vector multiplet can be computed in the same way i.e. by first computing the single letter index and then taking the plethystic exponent. We have tabulated the letters of the vector multiplet with $\delta=0$ in table \ref{tab2}. 

\begin{ex}[]{ex3}
Compute the single letter index of the vector multiplet with respect to $\widetilde{\CQ}_{\dot +}$.
\end{ex}

\begin{table}[htbp]
\begin{centering}
\begin{tabular}{|c||c|c|c|c||c|c|}
\hline Letters & $\Delta$ & $j_{1}$ & $j_{2}$ & $r$ 
& $\delta$ & ${\cal I}$  \tabularnewline \hline \hline 
${\bar \lambda}_{\dot +}$ &
$\frac{3}{2}$ & $0$ & $\frac{1}{2}$ & $-1$ & 
$0$ & $-p$ \tabularnewline \hline
${\bar \lambda}_{\dot -}$ &
$\frac{3}{2}$ & $0$ & $-\frac{1}{2}$ & $-1$ & 
$0$ & $-q$ \tabularnewline \hline
$F_{++}$ & $2$ & $1$ & $0$ &
$0$ & $0$ & $pq$  \tabularnewline
\hline 
$\partial_{+}^{\dot \alpha} {\bar \lambda}_{\dot \alpha}$ & $\frac52$ & $\frac12$ & $0$ & $-1$ 
& $0$ & $pq$
 \tabularnewline \hline\hline
$\partial_{+{\dot +}}$ & $1$ & $\frac{1}{2}$ & $\frac{1}{2}$ &
$0$ &  $0$ & $p$ \tabularnewline \hline
$\partial_{-{\dot +}}$ & $1$ & $-\frac{1}{2}$ & $\frac{1}{2}$ &
$0$ &  $0$ & $q$ \tabularnewline \hline
\end{tabular}
\par\end{centering}
\caption{\label{tab2}
In this table we have only tabulated the letters with $\delta=0$. These are the letters that contribute to the index.
}
\end{table}

As a result the single letter index for the vector multiplet is
\be
i_V(a_i:p,q)=\frac{-p-q+2pq}{(1-p)(1-q)} \chi_{adj}(a_i)=\Big(-\frac{p}{1-p}-\frac{q}{1-q}\Big)(\sum_\alpha a^\alpha +N)
\ee
Here we have multiplied the single letter index by the character of the adjoint representation of the gauge group. This is because the vector multiplet is always in the adjoint representation. The character of the adjoint is $\chi_{adj}(a_i)=\sum_\alpha a^\alpha +N$ where $\alpha$ are the root vectors. The additive factor of $N$ corresponds to the Cartan generators which have zero weight vector.

Now that we know the index of the chiral multiplet and the vector multiplet, we are ready to compute the index of gauge theories. Depending on the matter content, the supersymmetric gauge theories at low energies flow to non-trivial superconformal field theories. Most likely, it is impossible to compute their spectrum exactly. The superconformal index however, is easy to compute. The trick is to realize that the index is invariant under the change in gauge coupling. The index of strongly coupled gauge theories can be computed by setting the gauge coupling to zero. 

This means we only want to enumerate words that can be constructed out of letters with $\delta=0$. The partition function over such words is the plethystic exponent of the single letter index. These are all the supersymmetric local operators that can be constructed in four dimensions\footnote{In three dimension and five dimension, there are additional local operators that can not be constructed using letters of basic fields appearing in the Lagrangian. In three dimension they are known as the ``monopole operators" and in five dimension they are ``instanton operators". Accounting for them implies summing over sectors with nontrivial fluxes over $S^2$ and $S^4$ respectively. In these lectures we will only focus on four dimensional theories.}. However, in gauge theories we are only interested in constructing operators that are invariant under gauge transformations. These are operators that are singlet i.e. transform under trivial representation under the gauge group. 

Hence, the index of the gauge theory is
\be
\CI(b_k;p,q)= \frac{1}{|W|}\oint \Big(\prod_{i=1}^N \frac{d a_i}{2\pi i a_i} \Big) \Delta(a_i) \, \CI_V(a_i;p,q) \prod_{\phi_i} \CI_{\phi_i} (a_i,b_k;p,q).
\ee
Here $a_i$ are the fugacities for the gauge symmetry and $b_k$ are the fugacities for all other symmetries (global symmetries) that the chiral multiplets may transform under. 

Using the fact that $\Delta(a_i)=(PE[\sum_\alpha a^\alpha])^{-1}$, we can simplify the combination,
\bea
\Delta(a_i) \, \CI_V(a_i;p,q)&=&PE[\Big(\Big(-\frac{p}{1-p}-\frac{q}{1-q}\Big)\chi_{adj}(a_i)-\sum_\alpha a^\alpha\Big) ]\\
&=& PE[\Big(\frac{pq-1}{(1-p)(1-q)}(\sum_\alpha a^\alpha)+N \Big(-\frac{p}{1-p}-\frac{q}{1-q}\Big)\Big)]\nonumber\\
&=&\kappa^N\prod_{\alpha} \Gamma(pq a^\alpha),\qquad \kappa\equiv (p,p)(q,q), \quad(a,q)\equiv \prod_{i=0}^\infty (1-a q^i)\nonumber
\eea
The function $(a,q)$ is called the Pochhammer symbol. We have used a shorthand $\kappa$ to denote the index of a single $U(1)$ vector multiplet.

Using this simplification the gauge theory index can be cast completely in terms of the Elliptic gamma functions (and the Pochhammer symbol).
\bea\label{general-gauge}
\CI(b_k;p,q)&=& \frac{\kappa^N}{|W|}\nonumber\\
&\times &\oint \frac{d a_i}{(2\pi i)^N}\Gamma(pq a^\alpha) \prod_{\phi_i} \prod_{\rho\in R_i^G , {\rho'} \in {R'}_i^F}\Gamma((pq)^{\frac12 r_i} a^\rho b^{\rho'}).
\eea
Here we have assigned an arbitrary R-charge $r_i$ to the chiral multiplets. As discussed previously, it is fixed using superpotential and in addition possibly a-maximization.

\subsection*{Example}

Now we give an example of an index computation in $\CN=1$ gauge theory that illustrates the power of the superconformal index. Consider ${\cal N}=1$ $SU(2)$ gauge theory with three flavors of fundamental and anti-fundamental quarks. This theory has $SU(3)\times SU(3)\times U(1)$ global symmetries. Due to presence of the abelian factor, we need to compute the right R-charges using a-maximization. 
They turn out to be $\frac13$. Using all the general formula \eqref{general-gauge} for the index of gauge theories, we write the index of this gauge theory as the integral, 
\be\label{sugau}
{\cal I} = \kappa\oint \frac{dz}{4\pi i z} \frac1{\Gamma(z^{\pm2})} \prod_{i=1}^3
\Gamma((pq)^{\frac16} b u_i z^{\pm1})\Gamma((pq)^{\frac16} b^{-1} v_i z^{\pm1})\,.
\ee 
Here $\prod_{i=1}^3u_i=\prod_{i=1}^3 v_i=1$, with these fugacities parametrizing the $SU(3)_u\times SU(3)_v$ flavor symmetry rotating the fundamental and anti-fundamental quarks, while $b$ parametrizes the baryonic $U(1)_b$. Usually, as physicists we would deal with these integrals by setting $p=xy, q=x/y$ and performing the integral in a Taylor series in $x$ but this particular integral has been studied in the math literature. Remarkably it admits a closed form expression,
\be\label{spiri} 
{\cal I} = \prod_{i<j} \Gamma((pq)^{\frac13} t_i t_j)\,. 
\ee 
Here $\{t_i\}= \{ b u_i,\, b^{-1} v_i\}$. This is known as Spiridonov's beta function identity \cite{Spiridonov_2001}.

\begin{ex}[]{ex4}
Verify the identity \eqref{spiri} in Taylor series expansion in $x$.
\end{ex}

This form is extremely suggestive. It is the index of a theory with fifteen chiral multiplets interacting with a superpotential that imparts each of them an R-charge of $2/3$. Moreover, that the $u$ and $v$ fugacities combine to form six $t_i$ such that the index is invariant under their permutation suggests that this theory has $SU(6)$ symmetry whose fugacities are $t_i$ (The symmetry is $SU(6)$ rather than $U(6)$ because $\prod_i t_i=1$). Such a superpotential is simply a Pfaffian of the $6\times 6$ anti-symmetric matrix one constructs out of these fifteen chiral fields. 

This suggest a duality between the gauge theory and theory of only chiral multiplets. This is indeed true. This is one of  the many celebrated Seiberg's dualities \cite{Seiberg:1994pq}. It just so happens that the consequence of this particular duality on the superconformal index has been studied in the mathematical literature independently.  But more generally,  in this way, all Seiberg dualities become source of non-trivial mathematical integral identities\footnote{The superconformal indices of these theories at large $N$ were studied in \cite{Dolan:2008qi}}. In fact the integrals of these types, called the elliptic hypergeometric integrals, have been studied separately by mathematicians. See \cite{Spiridonov:2019dov} for an introduction and references. More importantly from a physicists point of view,  matching of superconformal index provides a non-trivial check on remarkable non-perturbative duality conjectures.

\subsection{Invariance under Recombination}
In supersymmetric quantum mechanics, the Hilbert space forms a representation of the supersymmetry algebra $SU(1|1)$. There are two types of irreducible representations. A two dimensional \emph{long multiplet} $\{|\chi\rangle, |\psi\rangle\}$
\be
Q |\chi\rangle=|\psi\rangle, \qquad Q|\psi\rangle =0
\ee
and a one dimensional \emph{short multiplet} $\{|\chi\rangle\}$
\be
Q|\phi\rangle=0.
\ee
Just from the algebra $\{Q,Q^\dagger\}=H$, it is easy to see that $H|\phi\rangle=0$. If we make a smooth supersymmetric  deformation of the theory, energy of a long multiplet may hit zero and at that point it will split into two short multiplets. Alternatively, if there are two short multiplets at zero energy, then a smooth supersymmetric deformation of the theory will generically cause them to \emph{recombine} to form a long multiplet. The crucial property of the Witten index is that it is invariant under recombination which is why it is protected under smooth supersymmetric deformation.

One can ask the same representation theoretic question about the superconformal index and understand it as a quantity that is invariant under all the recombinations allowed by the superconformal algebra. Lorenz Eberhardt  has already talked about superconformal algebras and their representations, long and short. We will draw on that discussion for this purpose. Although recombination rules have been first discussed in \cite{Dolan:2002zh}, we will follow the exposition of \cite{Kinney:2005ej}. 
Let us recall, 

The three types of shortening conditions
\begin{align}
& A_1: \epsilon^{\alpha \beta}\CQ_\alpha ([j_1,j_2]_{\Delta}^r)_\beta=0\quad  \Rightarrow \quad \Delta=2+2j_1-\frac32r\qquad \text{Null state:}\qquad [j_1-\frac12,j_2]_{\Delta+\frac12}^{r-1} \nonumber\\
& A_2: \epsilon^{\alpha \beta}\CQ_\alpha\CQ_\beta [0,j_2]_{\Delta}^r=0\quad  \,\,\Rightarrow \quad \Delta=2-\frac32r\qquad\qquad \,\,\,\text{Null state:}\qquad [0,j_2]_{\Delta+1}^{r-2} \nonumber\\
& B_1: \CQ_\alpha [0,j_2]_{\Delta}^r=0\qquad\qquad  \Rightarrow \quad \Delta=-\frac32r\qquad\qquad \quad\,\,\,\text{Null state:}\qquad [\frac12,j_2]_{\Delta+\frac12}^{r-1} 
\end{align}
We have used the conventions of \cite{Cordova:2016emh} for classifying the shortening conditions. 
Consider the $A_1$ type shortening condition, when the primary of the superconformal multiplet saturates the unitarity bound $\Delta=2+2j_1-\frac32r$, the state $[j_1-\frac12,j_2]_{\Delta+\frac12}^{r-1}$ becomes null. Interestingly, the null state is also at the unitarity bound as can be checked easily. If you now think of the multiplet with the null state being a primary, that multiplet also has a null state and again at the unitarity bound. In this process, the value of $j_1$ is decreasing and the chain stops at $j_1=0$. In that case, the null state satisfies the $B_1$ type shortening condition. All in all, we have a chain of short multiplets such that any two consecutive multiplets can recombine to form long multiplets,
\be
[j_1,j_2]_\Delta^{r} \rightarrow [j_1-\frac12, j_2]_{\Delta+\frac12}^{r-1}\rightarrow \ldots \rightarrow [0,j_2]^{r-2j_1}_{\Delta+j_1}\rightarrow [0,j_2]_{\Delta+j_1+1}^{r-2j-2}.
\ee
What happens at the two ends of the chain? Clearly, the chain can not continue arbitrarily to the right as conformal dimension is decreasing in that direction. On the other hand, the $B_1$ type multiplet at the right end has the null state $[\frac12, j_2]_{\Delta+j+\frac12}^{r-2j-3}$. This null state has $j_1\neq 0$ and it is quick to check that this state is below the unitarity bound, so there is no question of considering a multiplet based on this primary. In this way, the recombination chain terminates in both directions.  For the purpose of recombinations, it is convenient to think of the $B_1$ type multiplet at the end as an $A_1$ type multiplet but with $j_1=-\frac12$ i.e.
\be
[0,j_2]_{\Delta=-\frac32r}^{r}\equiv [-\frac12,j_2]_{\Delta=-\frac32r-\frac12}^{r+1}
\ee

The superconformal index needs to capture all the spectral information modulo recombination and as any two consecutive multiplets in above chain can recombine, the index should count
\be
N_{r'=r-2j_1,j_2}=\sum_{p=-1}^{k} \,(-1)^p\,n([\frac{p}{2},j_2]_{\Delta=2+p-\frac32(r'+p)}^{r'+p})
\ee
Indeed, the index of the $A_1$ type multiplet $[j_1,j_2]_\Delta^{r}$ is exactly \cite{Gadde:2010en},
\be
\CI_{[j_1,j_2]_\Delta^{r}}=(-)^{2(j_1+j_2)+1}\frac{(pq)^{\frac12(-r+2j_1+2)}\chi_{j_2}(\sqrt{\frac{p}{q}})}{(1-p)(1-q)}.
\ee
If we evaluate this on a collection of multiplets in the recombination chain, it precisely counts $N_{r'=r-2j_1,j_2}$. This is because, the index is the same for all the multiplets in the chain except for the sign which alternates.

The $\CN=1$ superconformal algebra is $SU(2,2|1)$, the commutant of a single supercharge inside this algebra is $SU(2,1)$. The index must be a character of this algebra. Indeed, the two Cartan generators of this algebra as embedded in $SU(2,2|1)$ are $r-2j_1$ and $j_2$. This is as expected from the recombination rules.

\subsection{Intermission: overview of supersymmetric Lagrangians}
\subsection*{$\CN=0$ Lagrangians:}
A rule of thumb is that more the supersymmetry less is the data required to specify the Lagrangian. For example, if one wants to construct a general Lagrangian of scalars, fermions and gauge fields. First one would have to specify the charges (or more generally representations) of the matter fields under the gauge symmetry. In addition to the gauge interactions, the matter could have relevant interactions of its own such as Yukawa and quartic scalar in a way so as to preserve the gauge symmetry. Finally, the number of fermions and their representations need to be carefully chosen so that the gauge symmetry is non-anomalous.

\subsection*{$\CN=1$ Lagrangians:}

With $\CN=1$ supersymmetry, the classification simplifies somewhat (although it is nowhere close to being as simple as with $\CN=2$). Scalars and fermions are in one multiplet, the chiral multiplet and gauge fields and gaugino fermions are in one multiplet, vector multiplet. Specifying the representations of the charges of the chiral multiplet under the gauge symmetry totally fixes the chiral-vector interaction. For $\CN=0$, the matter interactions were of two types and were independent of each other, but with $\CN=1$ supersymmetry they get related to each other and can be written as a superpotential $W(\Phi_i)$. Again, one needs to be ensure the matter content is such that the gauge symmetry is non-anomalous.

\subsection*{$\CN=2$ Lagrangians:}
$\CN=2$ Lagrangians are constructed using an $\CN=2$ hypermultiplet, which consists of a pair of chiral multiplets, and an $\CN=2$ vector-multiplet, which consists of $\CN=1$ vector multiplet and $\CN=1$ chiral multiplet both transforming under the adjoint representation of the gauge group.
The Lagrangians with $\CN=2$ supersymmetry are much more easily specified than those with $\CN=1$. One only needs to specify the coupling of the vector-matter, the coupling between matter-matter is prohibited by $\CN=2$ supersymmetry. The former coupling is simply given by specifying the representation of the hypermultiplet under the gauge group. This allows for a convenient graphical notation called quivers to describe $\CN=2$ theories. Moreover, as the matter comes in a pair of conjugate representations, the gauge symmetry of an $\CN=2$ theory is always non-anomalous. So a Lagrangian specified in this way is always a good Lagrangian (however it may not be asymptotically free). When the one-loop beta function of the gauge coupling vanishes, the gauge coupling  serves as an exactly marginal deformation. In that the case, the interacting superconformal theory parametrized by the complexified gauge coupling.

The superconformal index of $\CN=2$ theories can be computed by thinking of them as $\CN=1$ theory. But importantly, the commutant of $\CN=1$ algebra inside $\CN=2$ contains an abelian charge, and the fugacity for that charge can also be turned on. In this way, the index becomes a function of three variables (apart from global symmetry fugacities).

\subsection*{$\CN=3$ Lagrangians:}
There are no $\CN=3$ Lagrangians which are not $\CN=4$ symmetric. Although, in recent years a few non-Lagrangian $\CN=3$ (but not $\CN=4$) theories have been discovered \cite{Aharony:2015oyb, Garcia-Etxebarria:2015wns, Aharony:2016kai}. The commutant of $\CN=1$ inside $\CN=3$ is rank two. Hence two additional fugacities can be turned on, making the index a function of four variables. See \cite{Bourton:2018jwb} for the discussion of index of $\CN=3$ theories.

\subsection*{$\CN=4$ Lagrangians:}
The $\CN=4$ Lagrangian is constructed with a single vector multiplet in the adjoint representation. It is always conformal with complexified gauge coupling parametrizing the space of superconformal theories. The $\CN=4$ vector multiplet consists of $\CN=2$ vector multiplet and $\CN=2$ hypermultiplet in the adjoint representation. In terms of $\CN=1$ multiplets, it consists of an $\CN=1$ vector multiplet and three $\CN=1$ chiral multiplets that transform in the adjoint. 
The commutant of $\CN=1$ inside $\CN=4$ is of rank two\footnote{The reason that it is not three as one would have naively guessed is because the R-symmetry is $SU(4)$ and not $SU(4) \times U(1)$. For $\CN=1,2,3$, the R-symmetry is $SU(\CN)\times U(1)$.}. So two more fugacities can be turned on, making the index a function of four variables.

If the superconformal algebra has rank $r$ then the most general index is a function of $r-2$ fugacities. This is because the index only counts states with $\delta=0$ and we are only allowed to turn on fugacities for charges that commute with $\CQ$. These two conditions reduce the number of variables by $2$.

\section{$\CN=2$ theories}\label{n=2index}
The $\CN=2$ superconformal algebra consists of two supercharges $\{ {\cal Q}_{I\alpha} \, , {\cal S}^{I\alpha} \equiv {\cal Q}^{\dagger\, I\alpha} \, , {\widetilde {\cal Q}^I_{\dot \alpha}} \, ,
\widetilde {\cal S}_I^{\dot \alpha} \equiv {\widetilde {\cal Q}_I^{\dagger \,\dot \alpha}} \}$ for $I=1,2$. In addition to the conformal algebra, the bosonic subalgebra consists of R-symmetry $SU(2)_R\times U(1)_r$. The supercharges ${\cal Q}_{I\alpha}$ transform as a doublet under $SU(2)$, explicitly denoted by indices $I$. Under $U(1)_r$,  ${\cal Q}_{I\alpha}$ has charge $-1$ and $\widetilde{\cal Q}^I_{ \dot \alpha}$ has charge $1$. The indices $\alpha$ and $\dot \alpha$ are respectively $SU(2)_1$ and $SU(2)_2$ indices, with  $SU(2)_1 \times SU(2)_2 = Spin(4)$ the rotational subgroup of the conformal group. All the $\CQ$'s and ${\widetilde \CQ}$'s have conformal dimension $1/2$ and all the  $\CS$'s and ${\widetilde \CS}$'s have conformal dimension $-1/2$.
The most important (anti)-commutation relation in superconformal algebras is 
\bea
\{\Q_{I\alpha}, \, {\cal \Q}^{\dagger\, J\beta} \} & =& \Delta \delta_\alpha^\beta \delta_I^J+2M_\alpha^\beta \delta_I^J-2 R_I^J \delta_\alpha^\beta+\frac12 r\delta_\alpha^\beta \delta_I^J \\
\{\widetilde \Q^J_{\dot \alpha}\,, \, {\widetilde {\cal \Q}}_I^{\dagger \, \dot \beta} \} & =& \Delta \delta_\alpha^\beta \delta_I^J +2 \widetilde M_{\dot \alpha}^{\dot \beta} \delta_I^J-2 R_I^J \delta_{\dot \alpha}^{\dot \beta}-\frac12 r\delta_\alpha^\beta \delta_I^J \, ,
\eea
Without loss of generality we pick $\CQ_{1-}$ to compute the index. It is useful to note explicitly,
\be
\{\CQ_{1-},(\CQ_{1-})^\dagger\}=\Delta-2j_1-2R+\frac12 r.
\ee
With a choice of the supercharge $\CQ_{1-}$, an $\CN=1$ subalgebra is determined. It is generated by  $\{ {\cal Q}_{1\alpha} \, , {\cal S}^{1\alpha} \equiv {\cal Q}^{\dagger\, 1\alpha} \, , {\widetilde {\cal Q}^1_{\dot \alpha}} \, ,
\widetilde {\cal S}_1^{\dot \alpha} \equiv {\widetilde {\cal Q}_1^{\dagger \,\dot \alpha}} \}$. Comparing with equation \eqref{N1Q},
\be
r_{\CN=1}=\frac23(-2R+\frac12 r).
\ee
Also all the supercharges commute with $R+\frac12r$. As the only charges that have non-trivial commutations with the R-symmetry are supercharges, $R+\frac12r$ is the commutant with respect to the entire $\CN=1$ algebra. We define the index by turning on a fugacity with respect to this charge,
\be
\II(p,q,t) \equiv {\rm Tr} \, (-1)^F p^{\frac13(\Delta+j_1)+j_2} q^{\frac13(\Delta+j_1)-j_2} x^{(R+\frac12 r)}\, ,\qquad
\delta=\Delta-2j_1-2R+\frac12 r\, ,
\ee

\subsection*{Index of the hypermultiplet}
The $\CN=2$ hypermultiplet consists of a pair of $\CN=1$ chiral multiplets. One chiral multiplet along with the complex conjugate of the other transform as a doublet $q_I$ under $SU(2)_R$ symmetry and have the same representation under global symmetry. This means the two chiral multiplets in the hypermultiplet transform in conjugate representation of the global symmetry. Let us look at their contribution to the index.

The chiral multiplet $q_1$ has  $r_{\CN=1}= \frac23 (1+0)$ and so does the other chiral ${\bar q}_2$. Moreover, they both have the same charge under the commutant $R+\frac12 r=-\frac12$. As a result the single letter index of the hypermultiplet is
\bea
i_{H}(a_i;p,q,x)&=&\frac{(pq)^{\frac13}  x^{\frac12} \chi_R(a_i)-(pq)^{\frac23}  x^{-\frac12} \chi_{\bar R}(a_i)}{(1-p)(1-q)}+\frac{(pq)^{\frac13}  x^{\frac12} \chi_{\bar R}(a_i)-(pq)^{\frac23}  x^{-\frac12} \chi_{ R}(a_i)}{(1-p)(1-q)}\nonumber\\
i_{H}(a_i;p,q,x)&=&\frac{(pq)^{\frac13}  x^{\frac12}-(pq)^{\frac23}  x^{-\frac12} }{(1-p)(1-q)}\Big(\chi_R(a_i)+\chi_{\bar R}(a_i)\Big)
\eea
A single chiral multiplet $q_1$ is sometimes known as the half-hypermultiplet. One can understand the contribution of the hypermultiplet as that of a half-hypermultiplet but transforming in representation $R$ as well as $\bar R$.

\subsection*{Index of the vector multiplet}
The $\CN=2$ vector multiplet consists of an $\CN=1$ chiral multiplet $\phi$ and an $\CN=1$ vector multiplet $V$. The chiral multiplet is neutral under $SU(2)_R$ but has charge $2$ under $U(1)_r$. The effective $r_{\CN=1}=\frac23(0+\frac12)=\frac23$ and the charge under the commutant is $R+\frac12 r=1$. The $\CN=1$ vector multiplet, of course, can't be charged under any global symmetry. As a result, the index of the $\CN=2$ vector multiplet is,
\be
i_V(a_i;p,q,x)=\Big(\frac{(pq)^{\frac13}  x^{-1} -(pq)^{\frac23} x   }{(1-p)(1-q)}+\frac{-p-q+2pq}{(1-p)(1-q)}\Big)\chi_{adj}(a_i).
\ee

\subsection{Supersymmetric limits}
One can take limits of the superconformal index so that not all short multiplets but only certain special short multiplets contribute to the index. 
Supersymmetric limits for the $\CN=2$ index have been classified and named in \cite{Gadde:2011uv}.
For this analysis it is convenient to use variable $t$ by replacement $x\to (pq)^{-\frac23} t$. 
In terms of these variables, the index is
\be
\CI(p,q,t)={\rm Tr} \,(-1)^F\,p^{\frac13(\Delta+j_1)+j_2-\frac23(R+\frac12 r)}\,q^{\frac13(\Delta+j_1)-j_2-\frac23(R+\frac12 r)} t^{R+\frac12 r}.
\ee
Using $\delta=\Delta-2j_1-2R+\frac12 r=0$, this can be recast as
\bea\label{delta}
\CI(p,q,t)&=&{\rm Tr} \,(-1)^F\, p^{\frac12 (\Delta+2j_2-2R-\frac12 r)}\,q^{\frac12(\Delta-2j_2-2R-\frac12 r)} t^{R+\frac12 r},\nonumber\\
&=&{\rm Tr} \,(-1)^F\, p^{\frac12 \{\widetilde\CQ_{1{\dot +}},(\widetilde\CQ_{1{\dot +}})^\dagger\}}\,q^{\frac12 \{\widetilde\CQ_{1{\dot -}},(\widetilde\CQ_{1{\dot -}})^\dagger\}} t^{R+\frac12 r}
\eea
Then the indices of the half-hyper multiplet and the vector multiplet become
\be
i_{\frac12 H}(p,q,t)=\frac{\sqrt{t}-pq/\sqrt{t} }{(1-p)(1-q)},\qquad i_V(p,q,t)=\frac{pq/t -t }{(1-p)(1-q)}+\frac{-p-q+2pq}{(1-p)(1-q)}
\ee
The advantage of writing the index as in equation \eqref{delta} is that the charges that appear are manifestly non-negative. This allows for taking either $p$ or $q$ or both variables to zero. As we see below, these limits count operators with enhanced supersymmetry.

\begin{ex}[]{ex5}
Define $\delta_\alpha=\{\CQ_\alpha,(\CQ_\alpha)^\dagger\}$ and correspondingly ${\tilde \delta}_{\dot \alpha}$ and construct the index using the charges $\delta_\alpha,{\tilde \delta}_{\dot \alpha}$. The advantage of this choice of charges is that they are positive definite so the corresponding fugacities can be taken to zero.
\end{ex}

\subsection*{Macdonald index:}
This corresponds to the limit $p\to 0$. In this limit, states with $\{\widetilde\CQ_{1{\dot +}},(\widetilde\CQ_{1{\dot +}})^\dagger\}=0$ contribute. This means they are annihilated by  $\widetilde\CQ_{1{\dot +}}$ in addition to getting annihilated by $\CQ_{1-}$. 
In this limit the single letter indices become,
\be
i_{\frac12 H}(q,t)=\frac{\sqrt{t}}{1-q},\qquad i_V(q,t)=\frac{-t-q }{1-q}.
\ee

\subsection*{Hall-Littlewood index:}
This is the limit of the index $p\to 0,q\to 0$. In this limit, the operators that are annihilated by $\widetilde\CQ_{1{\dot +}}$ and $\widetilde\CQ_{1{\dot -}}$ in addition to getting annihilated by $\CQ_{1-}$. The conditions $\delta=0, \Delta+2j_2-2R-\frac12 r=0, \Delta-2j_2-2R-\frac12 r=0$ can be simplified to 
\be
j_2=0, \qquad j_1=r,\qquad \Delta=2R+r.
\ee
The single letter indices in this limit are,
\be
i_{\frac12 H}(t)=\sqrt{t},\qquad i_V(t)=-t.
\ee

\subsection*{Schur index:}

The Schur limit is the limit of the index $q=t$.
\be
\CI(p,q)={\rm Tr} (-1)^F \,p^{\frac12 (\Delta+2j_2-2R-\frac12 r)} \, q^{\frac12 (\Delta-2j_2-2R-\frac12 r)+(R+\frac12 r)}
\ee
 The charges that appear in the index are $\Delta+2j_2-2R-\frac12 r$ and $\Delta-2j_2+\frac12 r$. 
 By construction these charges  commute with $\CQ_{1-}$ because the object that we are computing is an index with respect to $\CQ_{1-}$. Interestingly, these charges also commute with  ${\widetilde \CQ}_{1{\dot +}}$. As a result, the index only receive contribution from states with  $\{{\widetilde \CQ}_{1{\dot +}},({\widetilde \CQ}_{1{\dot +}})^\dagger\}=0$. As a result, the index is automatically independent of $p$ and the trace formula simplifies to
 \be
 \CI(q)={\rm Tr}\, (-1)^F\, q^{\Delta-R}.
 \ee 
The single letter indices in this limit are,
\be
i_{\frac12 H}(q)=\frac{\sqrt{q}}{1-q},\qquad i_V(q)=\frac{-2q}{1-q}.
\ee

The supersymmetric limits allows one to probe extra supersymmetric sectors of the $\CN=2$ superconformal algebra.

\subsection{S-Duality of $\CN=2$ super QCD}
Nontrivial dualities of a large class of $\CN=2$ gauge theories, known as the class $\CS$,  were discovered by Gaiotto \cite{Gaiotto:2009we}. The simplest duality of this kind, already known from the work of Seiberg and Witten \cite{Seiberg:1994aj} is of $SU(2)$ gauge theory with eight half-hypermultiplets. These half-hypers transform as a vector representation of $SO(8)$. Let us just focus on the subgroup $ SO(8)\supset SO(4)\times SO(4)=SU(2)_a\times SU(2)_b\times SU(2)_c\times SU(2)_d$. The $SO(8)$ decomposes into the representations of $SU(2)^4$ as ${\bf 8}_v=(2_a\otimes 2_b)\oplus(2_c\otimes 2_d)$. The character of this representation is 
\be
(a+\frac1a)(b+\frac1b)+(c+\frac1c)(d+\frac1d).
\ee
With this information we can immediately write down the integral formula for the index.
The index of this theory is
\be\label{su2}
{\cal I} = \kappa\Gamma(\frac{pq}{t})\oint \frac{dz}{4\pi i z} \frac{\Gamma(\frac{pq}{t} z^{\pm2})}{\Gamma(z^{\pm2})} \Gamma(\sqrt{t} z^{\pm 1} a^{\pm 1}b^{\pm 1})\Gamma(\sqrt{t} z^{\pm 1} c^{\pm 1}d^{\pm 1}).
\ee
Here we have used shorthand, $\Gamma(z^{\pm})=\Gamma(z)\Gamma(z^{-1})$ etc..

This theory is known to enjoy a triality. If we change the vector representation of half-hypers to spinor representation or conjugate spinor representation then one gets the same theory but at different coupling. As the index is insensitive to coupling, we expect the index to be invariant under these changes of representation. The spinor and conjugate spinor decompose under $SU(2)^4$ as
\be
{\bf 8}_s=(2_a\otimes 2_c)\oplus(2_b\otimes 2_d),\qquad {\bf 8}_c=(2_a\otimes 2_d)\oplus(2_b\otimes 2_c).
\ee
This means that the index must be invariant under the swap $b\leftrightarrow c$ and $b\leftrightarrow d$. In addition to these non-trivial invariances, the index is easily seen to be invariant under trivial swaps $a\leftrightarrow b$ and $c\leftrightarrow d$.

In mathematics literature, these invariances were discovered \cite{van_de_Bult_2011} around the same time they were anticipated from physics \cite{Gadde:2009kb}. As physicists we can quickly check these invariances by replacing $p=xy, q=x/y$ and expanding in powers of $x$.

\begin{ex}[]{ex6}
Do this.
\end{ex}

\subsection{Class S dualities and TQFT}
From the work of Gaiotto, it is known that this theory is obtained by compactifying 6d \emph{rank-1} $(2,0)$ theory on a sphere with four punctures. The complex structure parametrizing this surface maps to the complexified gauge coupling. See Bruno LeFloch's lectures for a detailed discussion.
Moreover, each of the $SU(2)$ factors in the $SU(2)^4$ global symmetry corresponds to the a puncture.

In the degeneration limit, the four punctured sphere degenerates and splits up into three punctured spheres. According to Gaiotto, a three punctured sphere is half-hyper multiplet in the tri-fundamental representation with respect to the $SU(2)$'s at the three punctures. Sphere with three punctures does not have any moduli, consistent with the theory of free half-hypers not having any coupling. There are three distinct degeneration limits that one could take, bring close either $a,b$ or $a,c$ or $a,d$. In each degeneration limit, we get a weakly coupled theory but where the half-hypers are in the representation  ${\bf 8}_v, {\bf 8}_s$ and ${\bf 8}_c$ representation of the $SO(8)$ global symmetry respectively. Focusing only on the $SU(2)^4$ subgroup, the duality can be thought of as a crossing symmetry.

More generally, as the index is independent of the gauge coupling and hence of the complex structure of the Riemann surface, it must be computed by a topological field theory. Abstractly, a topological field theory can be specified by giving three point functions and two point functions i.e. the propagator. 

Let us review  how this works. 
We parametrize the index of a  three-punctured sphere
as $\mathcal{I}(a,b,c)$, where $a,b,c$ are fugacities for the Cartan of the three $SU(2)$ symmetries.
On the other hand we can easily write down the 
``propagator'' associated to a cylinder,
\be
\eta(a,b) = \Delta(a) {\cal I}_V(a)\,  \delta(a, b^{-1})\,,
\ee
where  $\Delta(a)$ is the Haar measure and ${\cal I}_V(a)$  the index of a vector multiplet, which is known explicitly.
The index of a generic theory of class ${\cal S}$ can be written in terms of the index of these elementary constituents.
As the simplest example, gluing two three-punctured spheres with one cylinder one obtains the
 index of a four-punctured sphere,
\bea \label{following}
{\cal I} ( a,b,c,d)
&=& 
\oint \frac{d z}{2\pi i } \oint \frac{d x}{2\pi i} \;
 \mathcal{I}_H(a,b,z)\, \eta(z,x)   \,
\mathcal{I}_H(z,c,d)\, \\
&=& \oint d z\;\Delta(z)\,\mathcal{I}_H(a,b,z)\,{\mathcal I}_V(z)\,
\mathcal{I}_H(z,c,d).
\eea 
If we expand the index in a convenient basis of  functions 
$\{ f^\a(a) \}$,
labeled by $SU(2)$ representations $\{ \a \}$,\footnote{
For theories of type $A$,   $\{ f^\a({\mathbf a})\}$ are symmetric functions of their arguments, which are fugacities dual to the Cartan generators of $SU(k)$. More generally, for theories of type $D$ and $E$,
$\{ f^\a({\mathbf a})\}$ are invariant under the appropriate Weyl group.
}
we can associate to each  three-punctured sphere ``structure constants'' $C_{\a  \beta \gamma}$
and to each propagator a metric $\eta^{\a \beta}$,
\bea
{\mathcal I}(a,b,c) & = &
\sum_{\alpha,\beta,\gamma} C_{\alpha\beta\gamma}
\,f^\alpha(a)\,f^\beta(b)\,f^\gamma(c)\, \\
\eta^{\alpha \beta}& = &   
\oint \frac{d a}{2\pi i } \oint \frac{d b}{2\pi i }\;
 \eta(a,b) \,f^\alpha(a)\,f^\beta(b)\,.
\eea
 Invariance of the index
under the different ways to decompose the surface is tantamount of saying
that $C_{\a  \beta \gamma}$ and  $\eta^{\a \beta}$ define a two-dimensional topological QFT.\footnote{We are using this term
somewhat loosely. As axiomatized by Atiyah,  a TQFT is understood to have a finite-dimensional state-space,
while in our case the state-space will be infinite-dimensional. The best-understood example of a $2d$ topological theory with an infinite-dimensional
state-space is the zero-area limit of $2d$ Yang-Mills theory~\cite{Witten:1991we, Witten:1992xu} (see {\it e.g.}~\cite{Cordes:1994fc} for a comprehensive review). Happily, the $2d$ topological
theory associated to the index turns out to be closely related to  
$2d$ Yang-Mills.}
The crucial property is associativity, 
\be\label{asso}
C_{\a\beta\zeta}{C^{\zeta}}_{\gamma\delta}=C_{\a\gamma\zeta}{C^{\zeta}}_{\beta\delta}\, ,
\ee where indices are raised  with the metric $\eta^{\a\beta}$ and lowered with the inverse metric $\eta_{\a \beta}$.

It is very natural to choose the complete set of functions $\{ f^\a(a) \}$ to be orthonormal under the measure that appears in the propagator,
\be \label{orthrelnatural}
\oint \frac{d a}{2\pi i} 
\; \Delta(a)\, {\cal I}_V(a)\,  f^\a(a)   f^\beta(a)   = \delta^{\alpha \beta}\,.
\ee
Then  
the metric $\eta^{\alpha \beta}$ is trivial, 
\be
\eta^{\alpha \beta}= \delta^{\alpha \beta} \, .
\ee
Condition (\ref{orthrelnatural}) still leaves
considerable freedom, as it is obeyed by infinitely many bases of functions related by orthogonal transformations. 
The real simplification arises if we can find
an {\it explicit} basis $\{ f^\a(a)\}$, such that the structure constants
are {\it diagonal},  
\be\label{Ndiag}
C_{\a\beta\gamma}
 \neq 0\quad\to\quad \a=\beta=\gamma\,.
\ee 
Associativity  (\ref{asso}) is then automatic. 
For structure constants satisfying    (\ref{asso}) 
one can always find a basis in which they are diagonal.  The challenge is to describe the basis in concrete form.

 In general the measure appearing in the propagator is complicated and no explicit set of orthonormal functions is available.
 We find it very useful to consider an ansatz 
 \be \label{clever}
 f^\a(a)= {\cal K}(a) P^\a(a)\, ,
 \ee
for some function ${\cal K}(a)$. Clearly, from (\ref{orthrelnatural}), the functions $\{ P^\a(a)\}$ are orthonormal under the new measure $ \hat \Delta(a)$,
\be \label{orthrel}
\oint \frac{d a}{2\pi i} \;
\hat \Delta(a)\,  P^\a(a)   P^\beta(a)   = \delta^{\alpha \beta}\,,\qquad \hat \Delta(a) \equiv {\mathcal I}_V(a)\, {\mathcal K}(a)^2\,\Delta(a)\,.
\ee
Recall that $\Delta(a)$ is the Haar measure.
The name of the game is to find a clever choice of ${\cal K}(a)$, for which   $\hat \Delta(a)$ is a simple known measure
and the orthonormal basis $\{ P^\a(a)\}$ an explicit set of functions such that (\ref{Ndiag}) holds.

Once the diagonal basis $\{ f^\alpha (a) \}$ and the structure constant $C_{\alpha \alpha \alpha}$ are known, one can easily
calculate the index of the SCFT associated to the genus ${\frak g}$ surface with $s$ punctures. Such a surface can be built
by gluing $2 {\frak g} - 2 + s$ three-punctured spheres, so we have
\be \label{Igs}
{\cal I}_{{\frak g}, s} (a_1,a_2,\ldots, a_s) = \sum_\alpha (C_{\alpha \alpha \alpha})^{2 {\frak g} - 2 + s} \,\prod_{I=1}^s f^\alpha (a_I)\,.
\ee

\subsection*{Schur limit}
It is simplest to recognize the basis of orthonormal functions in the Schur limit. In this limit,
\be
\CI_V(a)=PE[\frac{-2q}{1-q}(a^2+1+a^{-2})],\qquad \CI_H(a,b,c)=PE[\frac{\sqrt{q}}{1-q}(a+\frac1a)(b+\frac1b)(c+\frac1c)].
\ee
If we choose ${\cal K}(a)=1/\sqrt{\CI_V(a)}$, $\hat \Delta(a)=\Delta(a)$. The basis of polynomials that is orthonormal with respect to the standard Haar measure are characters, also known as the Schur polynomials. 
Indeed, the three point function is diagonal in this basis \cite{Gadde:2011ik} i.e.
\be
\CI_H(a,b,c)=\frac{C(q)}{\sqrt{\CI_V(a)\CI_V(b)\CI_V(c)}}\sum_\alpha \frac{1}{{\rm dim}_q \alpha} \chi_\alpha(a)\chi_\alpha(b)\chi_\alpha(c).
\ee
Because, the orthonormal polynomials associated with this limit of the index are Schur polynomials, the limit is called the Schur limit. 

In the Hall-Littlewood and Macdonald limit, the orthogonal polynomials that appear are Hall-Littlewood and Macdonald polynomials respectively. For the full index, $P_\alpha(a)$ are not polynomials but turn out to be complicated functions associated to Ruijsenaars-Schnider integrable model. This is the model which is at the top of the Calogero-Moser hierarchy.

\subsection*{Generalization to higher rank}
As remarked earlier, the $\CN=2$ $SU(2)$ gauge theory is obtained by compactifying \emph{rank 1} $(2,0)$ SCFT in six dimensions. A similar web of dualities is obtained if one compactifies higher rank $(2,0)$ theories on Riemann surfaces but unlike in the rank $1$ case, majority of these theories do not have a Lagrangian description. In particular, for rank $N$, the theory associated to the three punctured sphere is  known as the $T_N$ theory. Theory $T_2$ is simply the theory of free hypers but $T_{N>2}$ are non-trivial strongly coupled SCFTs. This makes it impossible to take the direct route to computation of the index. However, because the expressions for the topological three point function and two point function are in a form that allow a natural lift to higher rank, one obtains natural conjectures for the superconformal indices of the entire family of strongly coupled theories. Again, it is simplest to exemplify in the Schur limit. As the index of the $T_2$ theory is
\be
\CI_{T_2}(a,b,c)=\frac{C_2(q)}{\sqrt{\CI_V(a)\CI_V(b)\CI_V(c)}}\sum_\alpha \frac{1}{{\rm dim}_q \alpha} \chi_\alpha(a)\chi_\alpha(b)\chi_\alpha(c),
\ee
the lift of this form to general $N$ would be
\be
\CI_{T_N}(a_i,b_i,c_i)=\frac{C_N(q)}{\sqrt{\CI_V(a_i)\CI_V(b_i)\CI_V(c_i)}}\sum_R \frac{1}{{\rm dim}_q R} \chi_R(a_i)\chi_R(b_i)\chi_R(c_i),
\ee
where $R$ runs over all the representations of $SU(N)$ and $a_i,b_i$ and $c_i$ are fugacities associated with global symmetries $SU(N)_i$. 

Similar conjectures have been  made for Macdonald limit \cite{Gadde:2011ik} and even for the full index \cite{Gaiotto:2012xa}.

 \section{Index at large $N$}\label{largen}
 In this section we will compute the superconformal index of $\CN=4$ $SU(N)$ super Yang-Mills in the large $N$ limit. This is done by approximating the integral by a saddle point. Same saddle point techniques also work for large $N$ limits of other supersymmetric gauge theories \cite{Nakayama:2005mf, Nakayama:2006ur, Dolan:2008qi, Gadde:2010en} but we will only focus on the $\CN=4$ case here.
When decomposed into $\CN=2$ multiplets, the $\CN=4$ super Yang-Mills theory contains one hypermultiplet and a vector multiplet, both in adjoint representation. The single letter index of a single $\CN=1$ vector multiplet is
 \bea
 i(p,q,x,s)&=&\frac{-p-q+2pq+(pq)^\frac13 x^{-1}-(pq)^{\frac23} x}{(1-p)(1-q)}\\
 &+&\frac{(pq)^{\frac13}x^{\frac12}s-(pq)^{\frac23}x^{-\frac12}s^{-1} +(pq)^{\frac13}x^{\frac12}s^{-1}-(pq)^{\frac23}x^{-\frac12}s}{(1-p)(1-q)}\nonumber\\
 &=&\frac{(pq)^\frac13(s\sqrt{x}+\frac{\sqrt{x}}{s}+\frac1x)-(pq)^\frac23(\frac{1}{s\sqrt{x}}+\frac{s}{\sqrt{x}}+x)-p-q+2pq}{(1-p)(1-q)}\nonumber
 \eea 
 Here we have introduced the fugacity $s$ for the charge that commutes with the $\CN=2$ subalgebra of $\CN=4$. It transforms the two half-hyper multiplets with charge $\pm 1$. To get the index of $SU(N)$ $\CN=4$ vector multiplet, we need to multiply this index by $\chi_{adj}(a_i)$ where $a_i$ are the fugacities associated with the $SU(N)$ symmetry.
 Interestingly, the index of a single $\CN=4$ vector satisfies,
 \be\label{factor}
 1-i(p,q,x,s)=\frac{(1-(pq)^\frac13 s\sqrt{x})(1-(pq)^\frac13 \frac{\sqrt{x}}{s})(1-(pq)^\frac13 \frac1x)}{(1-p)(1-q)}.
 \ee
 This will be useful later. 
 
 The index of the gauge theory is
 \bea\label{n4gauge}
 \CI&=&\frac{1}{N!}\oint \Big(\prod_{j=1}^N  \frac{da_j}{2\pi i}\Big) \Delta(a_i) PE[i(p,q,x,s)\chi_{adj}(a_i)]\nonumber\\
 &=&\frac{1}{N!}\oint \Big(\prod_{j=1}^N  \frac{da_j}{2\pi i}\Big) \exp\Big(-\sum_{n=1}^\infty\frac1n(1-i(p^n,q^n,x^n,s^n))\sum_{i\neq j} \frac{a_i^n}{a_j^n}  \Big)
 \eea
 This type of matrix model is analyzed in \cite{Aharony:2003sx} as well as in \cite{Kinney:2005ej} in the large $N$ limit. We follow their analysis.
It is convenient to introduce density of eigenvalues $\rho(\theta)$ normalized such that $\int_{-\pi}^\pi d\theta\rho(\theta)=1$. The index becomes a functional integral over $\rho(\theta)$ subject to the fact that $\rho$ must be non-negative. The action of the functional integral is
\bea
\CI&=&\int [D\rho(\theta)] \exp\Big(N^2 \int d\theta_1 d\theta_2 \rho(\theta_1) V(\theta_1-\theta_2) \rho(\theta_2)\Big)\nonumber\\
V(\theta)&=& -\sum_{n=1}^\infty\frac1n(1-i(p^n,q^n,x^n,s^n)\cos(n\theta).
\eea
Nice thing about these variables is that due to the factor of $N^2$ in the action, at large $N$, this integral becomes a saddle point integral. We further simplify it using modes of density $\rho_n=\int d\theta \rho(\theta) e^{in\theta}$ and potential $V_n=\int d\theta V(\theta) \cos(n\theta)$.  Note that $\rho_{-n}=\rho_{n}^\dagger$. 
The functional integral becomes
\bea
\CI&=&\int [d\rho_n d\rho_n^\dagger] \exp\Big(\frac{N^2}{2\pi} \sum_{n=1}^\infty\frac1n V_n|\rho_n|^2\Big)\nonumber\\
V_n&=&\frac{2\pi }{n}(1-i(p^n,q^n,x^n,s^n).
\eea
From \eqref{factor}, we see that $V_n<0$ for all $n\geq 1$. This makes all modes $\rho_{n\geq 1}$ massive. The classical configuration is dominated by $\rho_0$. The action for the zero mode is simply $0$ and the integral is given by the one-loop determinant,
\bea
\CI&=&\prod_{n=1}^\infty \frac{1}{1-i(p^n,q^n,x^n,s^n)}\nonumber\\
&=& \prod_{n=1}^\infty \frac{(1-p^n)(1-q^n)}{(1-(pq)^\frac{n}{3} s^nx^{\frac{n}{2}})(1-(pq)^\frac{n}{3} s^{-n}x^{\frac{n}{2}})(1-(pq)^\frac{n}{3} x^{-n})}.
\eea
It is easy to see that this closed form of the index is actually a plethystic exponent of a relatively simple function. 
\bea
\CI&=&PE[i_G(p,q,x,s)]\nonumber\\
i_G(p,q,x,s)&=&\frac{(pq)^\frac13 s\sqrt{x}}{1-(pq)^\frac13 s\sqrt{x}}+\frac{(pq)^\frac13 \frac{\sqrt{x}}{s}}{1-(pq)^\frac13 \frac{\sqrt{x}}{s}}+\frac{(pq)^\frac13 \frac1x}{1-(pq)^\frac13 \frac1x}-\frac{p}{1-p}-\frac{q}{1-q}.
\eea

At large $N$, $SU(N)$ $\CN=4$ super Yang-Mills is dual to a weakly coupled type $IIB$ string theory in $AdS_5\times S^5$. In the dual theory, only the supergravity states are protected (this is because when the gauge theory becomes strongly couple, string modes become very massive). This means the index of $\CN=4$ SYM should compute the index of supergravity fields in $AdS_5\times S_5$. The plethystic logarithm of $\CI$ i.e. $i_G(p,q,x,s)$ is indeed the single particle index of supergravity in  $AdS_5\times S_5$ \cite{Kinney:2005ej}.

\subsection{What about black holes?}
The supergravity in $AdS_5\times S_5$ also admits supersymmetric Black holes. They have entropy of $\CO(N^2)$. As we are finding that the index is of $\CO(1)$ in the large $N$ limit, the large number of states of the supersymmetric black hole are somehow not visible to the index. The reason for this is that the bosonic and fermionic states of the black hole  appear in a way that leads to severe cancellations among a very large number of states.  If these cancellations could be avoided then we would find a saddle point with action of $\CO(N^2)$ rather than $0$. This leads to a natural question: will the index ever tell about the black hole states?

The answer to this question depends on the nature of cancellations between the bosonic and fermionic states. Two types of cancellations that we want to highlight are best illustrated with an example. Consider a single variable limit of the index $p=q=t^3, x=s=1$. The single letter index in this limit is 
\be
i(t)=1-\frac{(1-t^2)^3}{(1-t^3)^2}.
\ee
Let us consider a Taylor series expansion of the index of $SU(N)$ theory for some large $N$
\be
\CI(t)=\sum_n a_n t^n.
\ee
The severe cancellation between Bose-Fermi states could be extremely \emph{fine} such that each coefficient in this series is small of $\CO(1)$ and not of $\CO(N^2)$. In this case, the index will never be able to detect large degeneracy of states associated to black hole entropy. But another possibility is that the cancellation is \emph{coarse} in the sense that the coefficients of the series are large of $\CO(N^2)$ but oscillate rapidly in sign. Even in this case, the large degeneracies are difficult to capture as the index $\CI(t)$ will only see  averaged out $\CO(1)$ degeneracies. 
However, there is a simple trick to get around this problem. If we take $t\to -t$ ($t\in {\mathbb R}$) then this might partially avoid the Bose-Fermi cancellations and index itself will be of $\CO(N^2)$. If the oscillations in sign of $a_n$ are with period $2$ then $t\to -t$ will perfectly avoid all the cancellations. If they are with period $3$ then we may need to take $t\to t e^{2\pi i/3}$  ($t\in {\mathbb R}$) to better avoid cancellations and so on. The lesson from this discussion is, if the cancellations between Bose-Fermi  states are course then one could perhaps capture $\CO(N^2)$ growth of degeneracy by turning on non-trivial phases for the fugacities.

Recently this possibility has been explored in \cite{Choi:2018hmj}, in the so called ``Cardy limit" of the index.  For this analysis, it turns out to be more convenient to reformulate the $\CN=4$ index in a more ``symmetric" way as we describe below. 
The commuting bosonic generators of $\CN=4$ superconformal algebra consists of three R-symmetry generators $r_{i=1,2,3}$, two rotation generators $h_{i=1,2}$ and a conformal dimension $\Delta$. The supercharges are labeled by their charges under the Cartan generators $\CQ_{h_1,h_2}^{r_1,r_2,r_3}$. These charges take values $\pm\frac12$ such that the product of all signs is $+$. 
The superconformal index  with respect to $\CQ_{--}^{+++}$ is
\be\label{n4index}
\CI={\rm Tr} \, e^{-\sum_{i=1}^3 \rho_i r_i}e^{-\sum_{i=1}^2 \omega_i h_i},\qquad \rho_1+\rho_2+\rho_3-\omega_1-\omega_2=2\pi i
\ee
\begin{ex}[]{n=4index}
Note that in equation \eqref{n4index} we have not included the factor $(-1)^F$. Show that equation \eqref{n4index} nevertheless  defines a superconformal index.
\\\\
Hint: The periodicity of all the chemical potentials is $4\pi i$.
\end{ex}

The single letter index of $\CN=4$ vector multiplet is
\be
i=1-2\frac{\sinh \frac{\rho_1}{2}\cdot \sinh \frac{\rho_2}{2}\cdot \sinh \frac{\rho_3}{2}}{\sinh \frac{\omega_1}{2}\cdot \sinh \frac{\omega_2}{2}}.
\ee
 In these variables, the index of the $SU(N)$ gauge theory \eqref{n4gauge} becomes,
 \be\label{kimn4}
 \CI=\frac{1}{N!}\int_0^{2\pi} \prod_{j=1}^N d{\alpha_j} \exp\Big(-\sum_{n=1}^\infty\frac{2}{n} \frac{\sinh \frac{n\rho_1}{2}\cdot \sinh \frac{n\rho_2}{2}\cdot \sinh \frac{n\rho_3}{2}}{\sinh \frac{n\omega_1}{2}\cdot \sinh \frac{n\omega_2}{2}}\sum_{i\neq j}e^{in\alpha_{ij}}\Big)
\ee
Instead of taking the large $N$ limit, we take the ``Cardy limit" i.e. ${\rm Re}(\omega_i)\to 0^+, {\rm Im}(\omega_i)=0$. It is called Cardy limit because it is reminiscent of taking high-temperature limit except that in this context this temperature couples to angular momenta rather than energy. Note that in this limit, ${\rm Re} (\rho_i)\to 0^+$ but due to the constraint $\rho_1+\rho_2+\rho_3=\omega_1+\omega_2+2\pi i$,  ${\rm Im}(\rho_i)$ for all $i=1,2,3$ can be $\CO(1)$.
In this limit we replace $\sinh(n\omega_i/2)\to n\omega_i/2$. Equation \eqref{kimn4} becomes,
 \be
 \CI=\frac{1}{N!}\int_0^{2\pi} \prod_{j=1}^N d{\alpha_j} \exp\Big(-\frac{1}{\omega_1\omega_2}\sum_{n=1}^\infty\frac{2^3}{n^3} \Big(\sinh \frac{n\rho_1}{2}\cdot \sinh \frac{n\rho_2}{2}\cdot \sinh \frac{n\rho_3}{2}\Big)\sum_{i\neq j}e^{in\alpha_{ij}}\Big)
\ee
The exponent can further be simplified in terms of the trilogarithm ${\rm Li}_3(x)\equiv \sum_{n=1}^\infty x^n/n^3$.
\bea
&&\sum_{n=1}^\infty\frac{2^3}{n^3} \Big(\sinh \frac{n\rho_1}{2}\cdot \sinh \frac{n\rho_2}{2}\cdot \sinh \frac{n\rho_3}{2}\Big)\sum_{i\neq j}e^{in\alpha_{ij}}=\sum_{s_1,s_2,s_3=\pm1} \sum_{i\neq j} {\rm Li}_3(e^{\frac{s_i\rho_i}{2}+i\alpha_{ij}})\nonumber\\
&=&\sum_{s_1s_2s_3=1} \sum_{i\neq  j} \Big({\rm Li}_3(e^{\frac{s_i\rho_i}{2}+i\alpha_{ij}})-{\rm Li}_3(e^{-\frac{s_i\rho_i}{2}-i\alpha_{ij}})\Big)
\eea
The trilogarithm enjoys the following property,
\be
{\rm Li}_3(e^x)-{\rm Li}_3(e^{-x})=-\frac{x^3}{6}+\frac{\pi i x^2}{2}+\frac{\pi^2 x}{3},\qquad {\rm For}\quad (0<{\rm Im}(x)<2\pi, {\rm Re}(x)\geq 0).
\ee
For $2\pi  p<{\rm Im}(x)<2\pi (p+1) $, we need to shift $x\to 2\pi i p$ on the right hand side. After some massaging, one notices that the saddle point of the action functional is at $\alpha_i=\alpha$. The classical value of the action at this saddle point goes as
\be
\log(\CI)\sim N^2\frac{\rho_1\rho_2\rho_3}{\omega_1\omega_2}.
\ee
This is precisely the functional that was found in \cite{Hosseini:2017mds} to produce the correct entropy of charged BPS black holes in $AdS_5$.

The same problem has been tackled in another way in \cite{Benini:2018ywd} by applying methods of Bethe ansatz to the superconformal index. For numerical studies of the index at finite $N$, see \cite{Agarwal:2020zwm, Murthy:2020rbd}.

\acknowledgments
We would like to thank the theoretical physics group at DESY and especially Elli Pomoni for extraordinary hospitality. We would also like to acknowledge our debt to the people of India for their steady support to study the basic sciences. 

\bibliographystyle{JHEP}
\bibliography{indexrefs}

\end{document}